\documentclass[prr, twocolumn, groupedaddress, showpacs, preprintnumbers, amsmath, amssymb, aps, floatfix, superscriptaddress]{revtex4-2}

\usepackage{siunitx}
\usepackage{graphicx}
\usepackage{verbatim}
\usepackage{dcolumn}
\usepackage[ruled,vlined]{algorithm2e}
\usepackage{color}
\usepackage{xcolor}
\usepackage[colorlinks = true, linkcolor = blue, urlcolor  = blue, citecolor = blue, anchorcolor = blue]{hyperref}
\usepackage{orcidlink}

\begin{document}


\title{Combining Methods for Localization of Linear Focusing Errors}


\author{I.~Morozov\,\orcidlink{0000-0002-1821-7051}}
\email{i.a.morozov@inp.nsk.su}
\affiliation{Budker Institute of Nuclear Physics SB RAS, Novosibirsk 630090, Russia}
\affiliation{SRF Siberian Circular Photon Source "SKIF" Boreskov Institute of Catalysis SB RAS, Koltsovo 630559, Russia}
\affiliation{Novosibirsk State Technical University, Novosibirsk 630073, Russia}
\author{Yu.~Maltseva\,\orcidlink{0000-0003-2753-2552}}
\affiliation{Budker Institute of Nuclear Physics SB RAS, Novosibirsk 630090, Russia}
\affiliation{Novosibirsk State Technical University, Novosibirsk 630073, Russia}
\date{\today}




\begin{abstract}

The primary objective of accelerator tuning is to correct its linear optics.
This involves adjusting a large set of parameters, such as quadrupole lens gradients and alignment errors, as well as addressing various calibration errors in beam position monitors (BPMs) and other components.
These errors can significantly impact the accelerator performance.
Using the full set of parameters might reduce the correction efficiency.
Typically, it is advisable to prioritize the correction of large errors, as their localization can enhance the overall efficiency of the correction process.
The task of error localization involves identifying the locations of potential error sources, along with their type.
This paper explores and compares several commonly used error localization methods, using the SKIF model to test and evaluate their efficiency.
Employing multiple methods enables cross-validation of the results obtained from the error localization process.
The combined results from several localization methods improve the overall accuracy of error localization.

\end{abstract}


\maketitle


\section{Introduction}
\label{sec:introduction}


The linear optical model of an accelerator is characterized by a large number of different parameters.
These parameters include, for example, quadrupole lens gradients, alignment errors of accelerator elements and beam position monitors (BPMs), as well as their calibration errors.
While some parameters can be adjusted in real-time during accelerator tuning, others require a specific setup for modification.


Correction of linear optical model errors is required for accelerator operation optimization.
The use of the full set of knobs simultaneously might in general reduce the effectiveness of the correction.
For instance, solutions obtained from optimizing the Euclidean norm contain several non-zero parameters, even when only one parameter is non-zero.
Alternative approaches, like using different norms~\cite{compressed, brunton} or applying regularization techniques~\cite{regularization}, might circumvent this issue.
Additionally, the optimization speed is affected by the number of parameters utilized, and using a large set of knobs might not be feasible.
A practical strategy involves grouping parameters and optimizing them in a sequential manner.
Another approach is based on localizing areas where potential errors might occur.
Each area, the section between two BPMs, corresponds to only a subset of the optimization parameters.
The use of a reduced number of parameters potentially enables more targeted corrections with enhanced efficiency.
Consequently, the correction task is divided into a series of sequential smaller tasks.
A smaller number of parameters allows for the more effective application of various optimization methods, such as RCDS~\cite{rcds} or the robust simplex algorithm~\cite{simplex}.


This paper examines the action and phase jump method~\cite{cardona2009, cardona2017} as well as the Twiss parameters transport method~\cite{maltseva, cern} for localizing linear focusing errors.
These methods calculate observable quantities that show jumps at error locations.
However, due to the challenge in automatically processing these jumps, we implement modifications where observable quantities reveal peaks at possible error locations, though this comes with a compromise in localization accuracy.
Both methods use turn-by-turn (TbT) data from coherent transverse beam oscillations captured by BPMs.


Furthermore, this paper develops multiple error localization methods where errors manifest as peaks in various observable quantities.
These methods depend on estimating beam centroid momenta~\cite{berz, ipac2016, ndmap}, Twiss parameters~\cite{castro, nbpm, better_nbpm, rupac2021_twiss, package}, linear uncoupled and coupled invariants, and transport matrices between BPMs from TbT data.
The underlying concept for all methods involves calculating specific observable quantities using only a segment of the accelerator,  enabling comparisons either between different segments or with corresponding model values.


\begin{figure*}[!t]
    \centering
    \includegraphics[width=\linewidth,height=0.33\textheight]{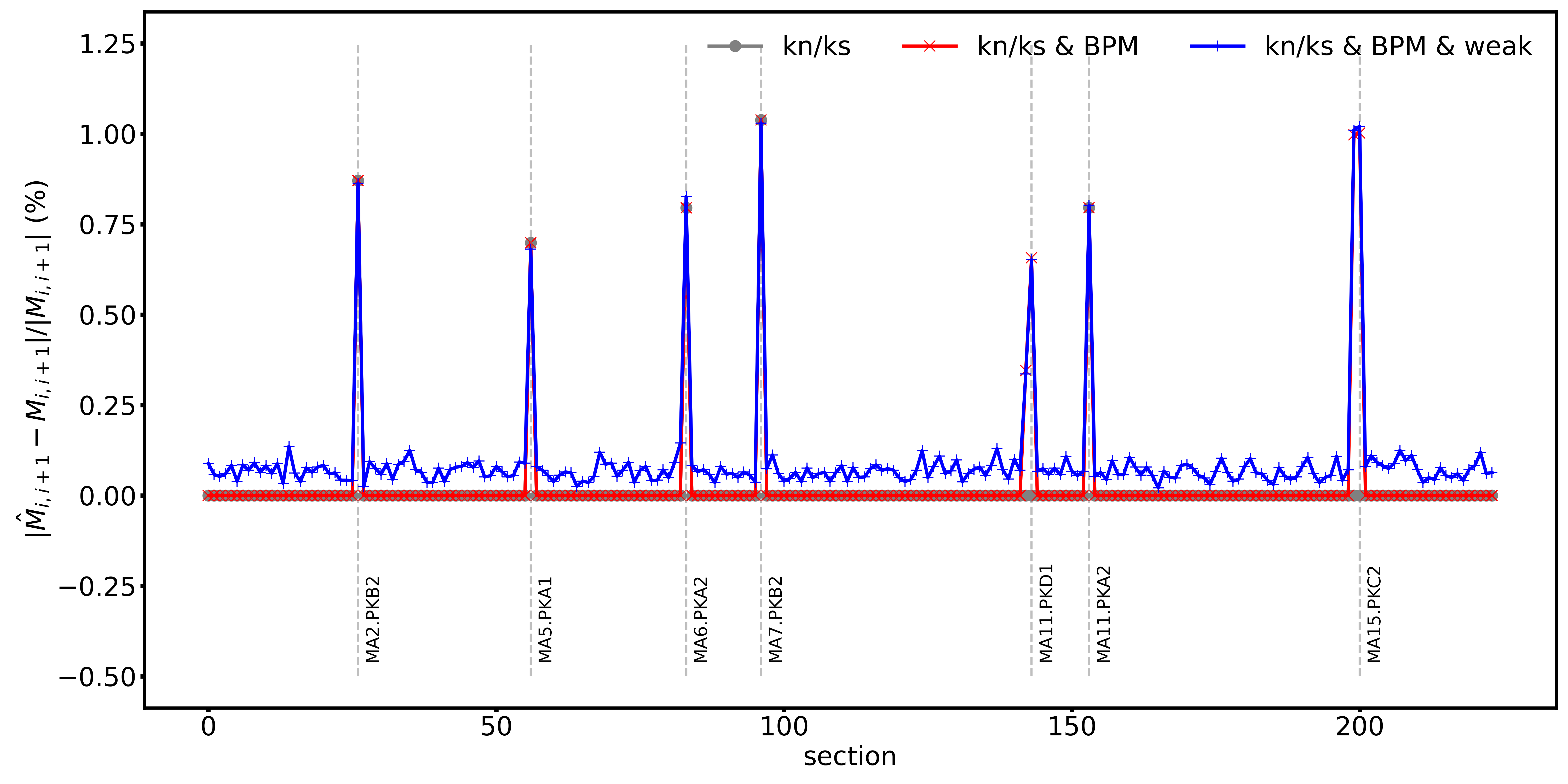}
    \caption{
    Errors in the norm of the transport matrices between adjacent monitors. Only strong quadrupole errors (in gray), all strong errors (in red), and all strong and weak errors (in blue).
    }
    \label{fig:error}
\end{figure*}


Since the localization is performed between BPMs, the cause of possible errors is due to elements within the section between BPMs or to BPMs themselves.
It should be noted that in the work~\cite{maltseva}, error localization is performed using neighboring quadrupole lenses.
However, such quadrupole lenses are expected to be separated only by a drift space.
When constructing observable quantities in each localization method, both the monitors and the sections between them are utilized.
Therefore, two outcomes are defined for each error localization method: one highlights the sections between monitors, and the other focuses on the BPMs themselves.
Linear optics measurements based on the phase of TbT signals~\cite{castro, nbpm, better_nbpm, rupac2021_twiss} are independent of BPM calibrations, allowing the identification of monitors with potential calibration errors.
Furthermore, localization methods can also be combined to enhance their overall localization accuracy.
This process involves normalizing the results from each method, followed by adding or multiplying these normalized indicators, similar to basic logical operations.
The combined results are also normalized, allowing for automatic identification of potential error sites based on peak locations.


The structure of this article is as follows:
Section~\ref{sec:setup} describes the test numerical problem and introduces various linear focusing errors into the SKIF storage ring lattice~\cite{skif}, along with the parameters of weak and strong errors.
Section~\ref{sec:common} details the action and phase jump method and the Twiss parameters transport method.
Forward differences represent peaks in the results of these methods, and the construction of error indicators for sections and monitors is explained. 
Sections \ref{sec:momenta}, \ref{sec:twiss}, \ref{sec:invariant}, and \ref{sec:matrix} introduce additional error indicators based on the estimation of momenta, Twiss parameters, linear invariants, and transport matrices between BPMs.
The procedure for combining different error indicators is presented in Section~\ref{sec:combine}.


\section{The Test Problem Setup}
\label{sec:setup}


To test the error localization methods, we use a simplified model of the SKIF storage ring~\cite{skif}. 
Several strong and weak errors are introduced into the lattice.
The model is represented by linear transverse transport matrices between several types of observation locations.
An error is considered strong if its effect leads to a significant change in the norm of the transport matrix of the section where the error is located.
Errors are added to all standalone quadrupole lenses, in the centers of long straights where insertion devices (IDs) are located, and in the BPMs.
An error in one or more elements of a given section affects only the transport matrix of that section, whereas a calibration error in a BPM affects two sections around it.


In standalone quadrupole lenses and IDs, thin focusing lenses are added as the source of errors, while for BPMs, calibration errors and longitudinal displacements are introduced.
All these errors are focusing errors that affect the observed linear optical model of the accelerator.
Sections between two adjacent monitors are named after the first monitor in each section.
The parameters for weak focusing errors can be found in Table~\ref{table:weak}.
These weak errors result in beta function beating ranging from \SIrange[range-phrase=\text{$\div$},range-units=single]{1}{3}{\percent}.
Random normal noise is also added to the simulated TbT signals obtained from BPMs.
The maximum oscillation amplitudes at BPM locations are around \SI{1}{mm}.
To calculate amplitudes and phases, only the first 128 and 256 turns are utilized, respectively.


\begin{table}[ht]
\begin{center}
\begin{tabular}{|l l|} 
 \hline
 Parameter & Value \\ [0.5ex] 
 \hline
 \hline
 $\sigma_{k_n} / \sigma_{k_s}$ & \SI{2.5E-4}{} \\ 
 \hline
 $\sigma_{n_x} / \sigma_{n_y}$ & \SIrange[range-phrase=\text{$\div$},range-units=single]{1}{2}{\micro\m} \\
 \hline
 $\sigma_{s_x} / \sigma_{s_y}$ & \SI{0.4E-3}{} \\
 \hline
\end{tabular}
\caption{\label{table:weak} Weak focusing errors parameters. $\sigma_{k_n} / \sigma_{k_s}$ -- $k_n$ and $k_s$ amplitude sigmas, $\sigma_{n_x} / \sigma_{n_y}$ -- monitor measurement noise sigmas, $\sigma_{s_x} / \sigma_{s_y}$ -- monitor scale calibration sigmas.}
\end{center}
\end{table}


The parameters for strong focusing errors can be found in Table~\ref{table:strong}.
Seven strong errors were introduced into the lattice: three in standalone quadrupole lenses, two in IDs, and two in BPMs.
Strong errors result in beta function beating at the level of \SI{15}{\percent}.
The magnitudes of strong errors are selected to ensure comparable effects on the norms of transport matrices from various error sources.
The impact of weak errors on these norms is about an order of magnitude smaller than that of strong errors.
Figure~\ref{fig:error} shows the norm errors in transport matrices between adjacent monitors under different combinations of focusing errors.


\begin{table}[ht]
\begin{center}
\begin{tabular}{|l l l l|} 
 \hline
 Name & Type & Parameter & Value \\ [0.5ex] 
 \hline
 \hline
 MA02.PKB2 & QUAD & $k_n / k_s$ & \SI{+5E-3}{} / \SI{-5E-3}{}\\ 
 \hline
 MA05.PKA1 & QUAD & $k_n$ & \SI{-5E-3}{} \\ 
 \hline
 MA06.PKA2 & ID & $k_s$ & \SI{-1E-2}{} \\ 
 \hline
 MA07.PKB2 & QUAD & $k_s$ & \SI{-1E-2}{} \\ 
 \hline
 MA11.PKD1 & BPM & $d_s$ & \SI{2}{\cm} \\ 
 \hline
 MA11.PKA2 & ID & $k_n$ & ~\SI{5E-3}{} \\ 
 \hline
 MA15.PKC2 & BPM & $g_x / g_y$ & \SI{0.985}{} / \SI{1.015}{} \\ 
 \hline
\end{tabular}
\caption{\label{table:strong} Focusing strong errors parameters. $k_n / k_s$ - thin focusing errors, $g_x / g_y$ - monitor scale calibration errors, $d_s$ - monitor longitudinal displacement errors.}
\end{center}
\end{table}


\section{Common Error Localization Methods}
\label{sec:common}


The method of action and phase jump is described in detail in \cite{cardona2009}.
This method is useful not just for localizing focusing errors but also for determining their type and magnitude, under certain assumptions.
In this paper, the method is used exclusively to identify sections of the accelerator where linear focusing errors occur.
The main idea of the method is to compute the initial action $I$ and phase $\varphi$ based on data from two adjacent beam position monitors. 
The accelerator lattice is considered as a transport channel with known model Twiss parameters.
This means that the transport matrices are known, allowing for the determination of the initial values  $I$ and $\varphi$. 
The initial conditions are expressed using coordinates from two adjacent monitors, $k$ and $k+1$, as follows
\begin{align}
    &I=\frac{\frac{{x_k}^2}{\beta_{k}}+\frac{x_{k+1}^2}{\beta_{k+1}}}{2{\sin(\varphi_{k+1}-\varphi_{k})}^2}-\frac{\frac{x_{k}x_{k+1}}{\sqrt{\beta_{k}\beta_{k+1}}}\cos(\varphi_{k+1}-\varphi_{k})}{\sin(\varphi_{k+1}-\varphi_{k})^2} \nonumber \\ 
    & \nonumber \\
    &\tan \varphi =\frac{\frac{x_{k}}{\sqrt{\beta_{k}}}\sin(\varphi_{k+1})-\frac{x_{k+1}}{\sqrt{\beta_{k+1}}}\sin(\varphi_{k})}{\frac{x_{k}}{\sqrt{\beta_{k}}}\cos(\varphi_{k+1})-\frac{x_{k+1}}{\sqrt{\beta_{k+1}}}\cos(\varphi_{k})} 
\end{align}
where $\beta_k$ and $\beta_{k+1}$ are the beta functions at the monitors, $\varphi_{k}$ and $\varphi_{k+1}$ are the phase advances from the reference point to each monitor, $x_k$ and $x_{k+1}$ are the transverse (horizontal or vertical) coordinate values.


\begin{figure}[!hb]
    \begin{center}
    \includegraphics[width=\columnwidth,height=0.45\columnwidth]{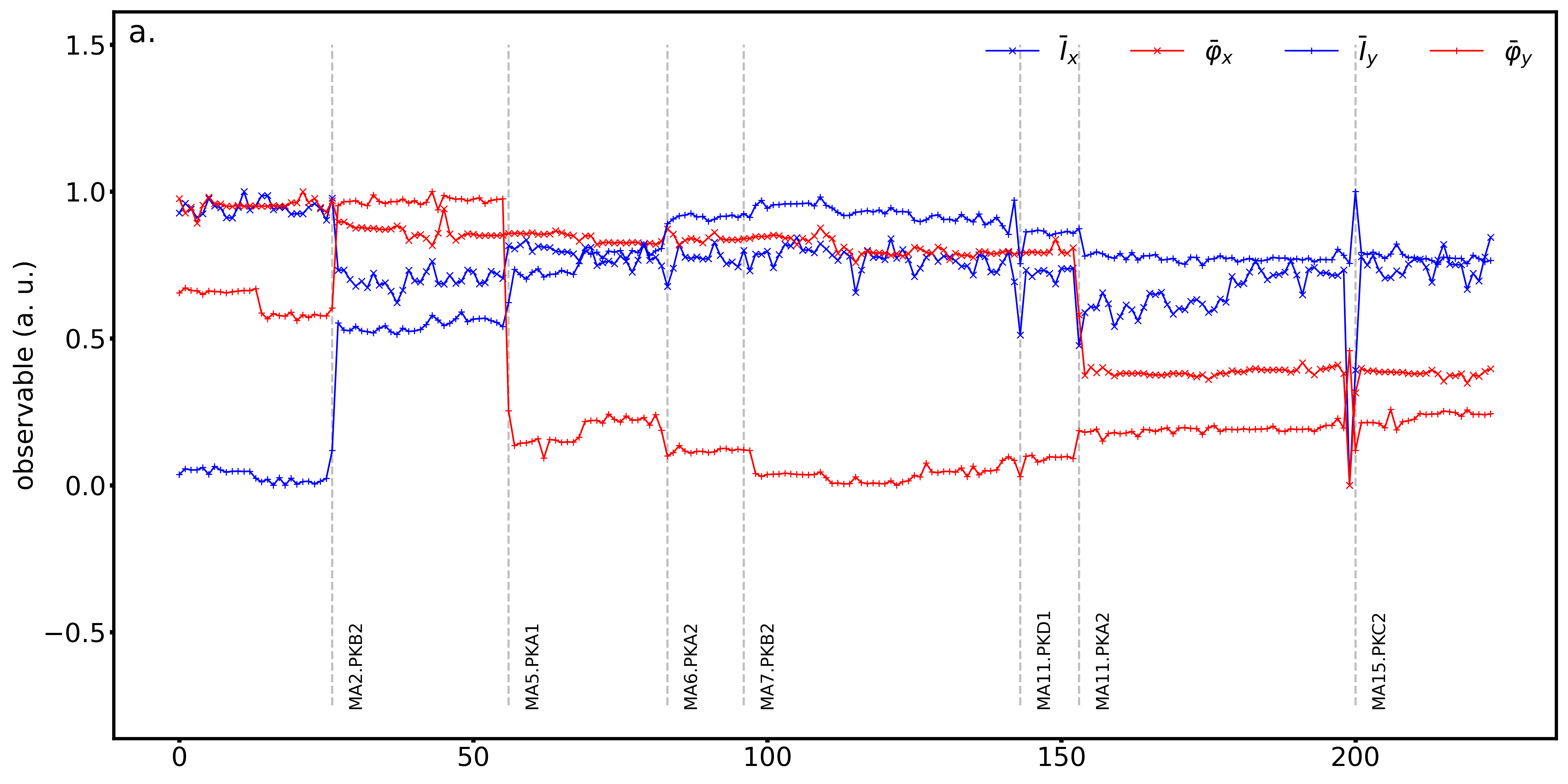}
    \includegraphics[width=\columnwidth,height=0.45\columnwidth]{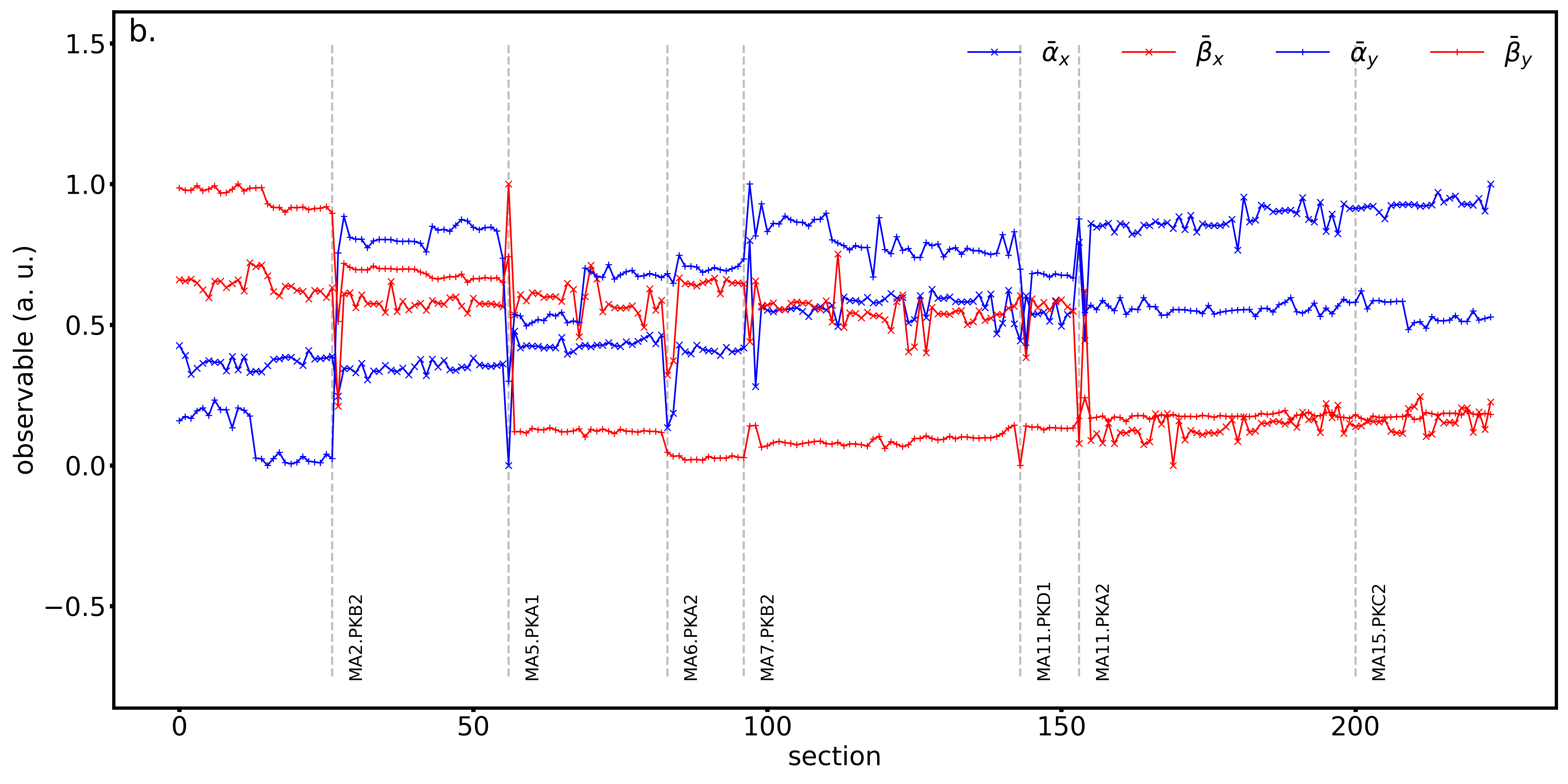}
    \end{center}
    \caption{
    Jumps in observable quantities for the action and phase jump method (a.) and the Twiss parameters propagation method (b.).
    For the (a.) case, each point corresponds to an interval between two adjacent monitors for which observable quantities are calculated.
    While for the (b.) case, each point corresponds to a single monitor, where Twiss parameters are calculated based on the two adjacent sections. Here parameters are computed at the common monitor located at the center of these two sections.
    }
    \label{fig:common}
\end{figure}


In the absence of errors, the values of $I$ and $\varphi$ will be the same for all sections.
However, errors cause noticeable jumps in these values.
The results of applying this method to the test problem are shown in Figure~\ref{fig:common}.
Here, TbT signals are used to compute $I$ and $\varphi$.
To reduce noise in TbT data, spatial modes from SVD decomposition of the beam history matrix at all monitors are utilized, rather than using the coordinate values themselves.
Calculation of each point on the graph requires using one section and two monitors.
For this specific test problem, the method shows low sensitivity to linear coupling errors.
Jumps due to linear coupling errors are hard to distinguish against the background of weak focusing errors.
However, as Figure~\ref{fig:error} shows, these errors significantly alter the norm of the corresponding transport matrices.
Notably, the method is sensitive to BPM scale calibrations, showing distinct jumps for pure calibration errors, indicating its potential in identifying these errors.


\begin{figure*}[!th]
    \begin{center}
    \minipage{0.5\linewidth}
      \includegraphics[width=\columnwidth,height=0.5\columnwidth]{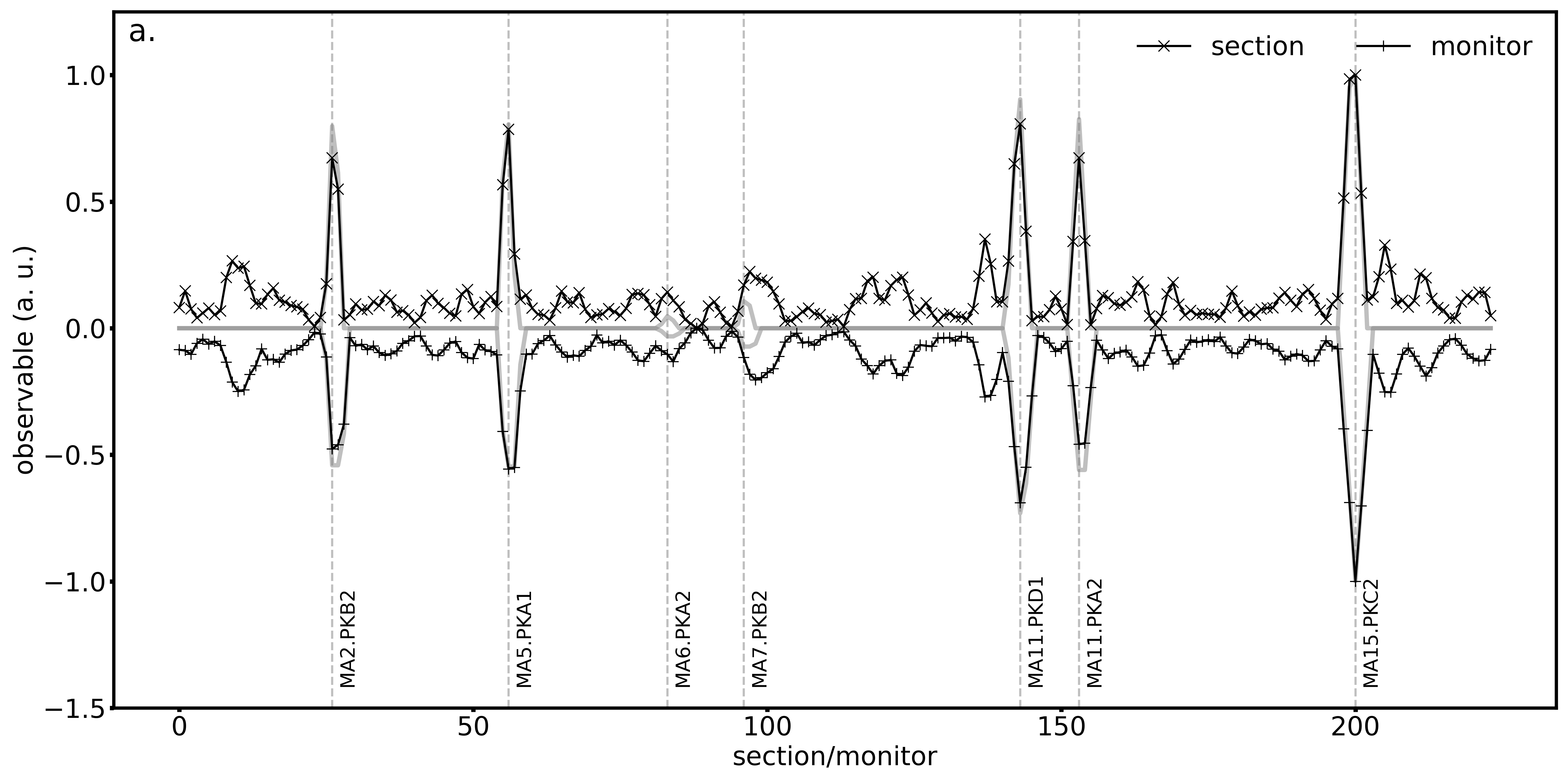}
    \endminipage
    \hfill
    \minipage{0.5\linewidth}
      \includegraphics[width=\columnwidth,height=0.5\columnwidth]{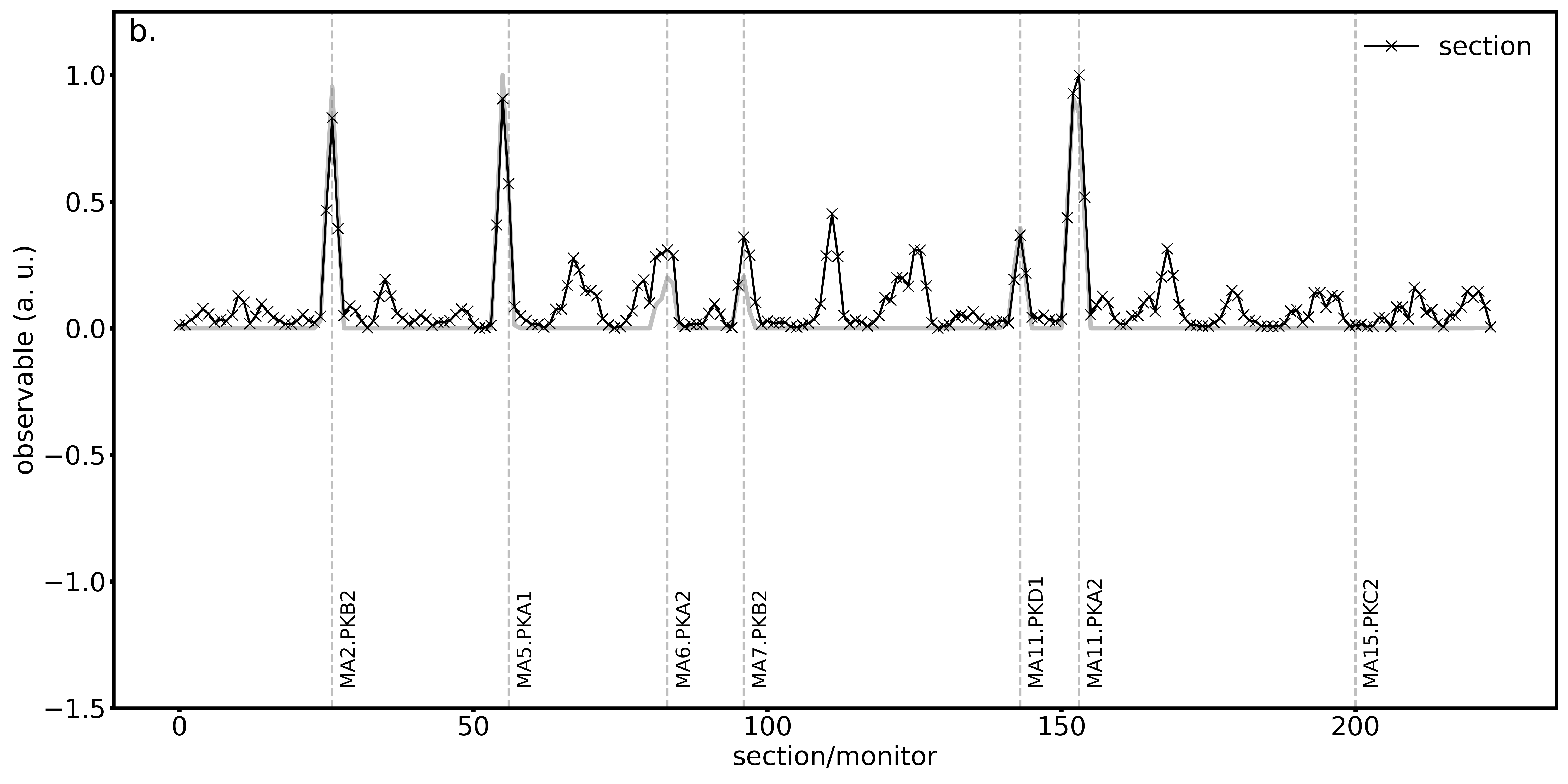}
    \endminipage
    \end{center}
    \caption{Error indicators based on the difference action and phase jump $2S3M$ (a.) and the difference Twiss parameters propagation $3S0M$ (b.). Results for monitors are reflected relative to the horizontal axis. Error indicators without weak errors are shown in gray.}
    \label{fig:mod}
\end{figure*}


Another error localization method is based on propagation of measured Twiss parameters~\cite{maltseva}.
Essentially, any method suitable for estimating a complete set of Twiss parameters can be applied.
Here, the method based on computation of the phases of TbT signals is used \cite{castro, nbpm, better_nbpm, rupac2021_twiss}. 
This method is capable of estimating all Twiss parameters, provided that coupling between transverse planes can be neglected.
A key advantage of this method is its independence from BPM scale calibration errors.
Thus, by comparing its results with those from the action and phase jump method, or any other method sensitive to calibration errors, one can identify potential calibration errors.
Estimation of Twiss parameters at a given BPM involves using the measured phases of TbT signals from that monitor and two adjacent monitors.
These uncoupled Twiss parameters are estimated using model values and the measured phase advances as follows
\begin{widetext}
\begin{align}
    &\alpha_i=\alpha^{m}_i\frac{\cot(\varphi_{ij})-\cot(\varphi_{ik})}{\cot(\varphi^{m}_{ij})-\cot(\varphi^{m}_{ik})} -\cot(\varphi_{ij}) \sec(\varphi^{m}_{ij} - \varphi^{m}_{ik})\cos(\varphi^{m}_{ik})\sin(\varphi^{m}_{ij}) + \cot(\varphi_{ik})\sec(\varphi^{m}_{ij}-\varphi^{m}_{ik})\cos(\varphi^{m}_{ij})\sin(\varphi^{m}_{ik}) \nonumber \\ 
    &\beta_i=\beta^{m}_i\frac{\cot(\varphi_{ij})-\cot(\varphi_{ik})}{\cot(\varphi^{m}_{ij})-\cot(\varphi^{m}_{ik})}
\end{align}
\vspace{-0.5cm}
\end{widetext}
where $\alpha^{m}_i$ and $\beta^{m}_i$ are the model Twiss parameters at the probed monitor $i$, $\varphi^{m}_{ij}$ and $\varphi^{m}_{ik}$ are the model phase advances from monitor $i$ to $j$ and $k$ respectively, while $\varphi_{ij}$ and $\varphi_{ik}$ are the corresponding measured phase advances.


Given uncoupled Twiss parameters at location $i$, they can be propagated to any other location $j$ using the transport matrix $M_{i,j}$
\begin{equation}
    {\begin{pmatrix} \beta & -\alpha & \\ -\alpha & \gamma &\end{pmatrix}}_j = 
    M_{i,j} {\begin{pmatrix} \beta & -\alpha & \\ -\alpha & \gamma &\end{pmatrix}}_i M_{i,j}^{T} 
\end{equation}


To increase phase estimation accuracy, TbT data are first examined for anomalies~\cite{rupac2021_anomaly}, then filtered using SVD decomposition~\cite{rupac2021_twiss}. 
In the estimation of Twiss parameters, errors in the relevant sections result in noticeable jumps in the propagated Twiss parameters.
Figure~\ref{fig:common} illustrates the application of this method to the test problem.
As expected, the method is insensitive to calibration errors but sensitive to even weak gradient errors, if the beta function is large at the location of these error sources.


From Figure~\ref{fig:common}, one can conclude that both methods successfully localized most of the errors.
The main disadvantage of the action and phase jump method is its relatively low sensitivity to linear coupling errors, potentially due to the nature of the test problem.
The Twiss parameters propagation method shows relative sensitivity to weak errors at locations with a large beta function.
To facilitate automatic processing and clearer distinction between monitor calibration errors and section errors, a transition to analysis of peaks is proposed.
This involves utilizing forward differences between adjacent results, thereby generating new observable quantities that are subsequently normalized.
These normalized quantities contain peaks at locations with linear focusing errors.
The methods differ in the number of sections ($*$) and monitors ($\#$) used for computing each observable quantity, denoted as $*S\#M$.
For example, the action and phase jump method is labeled as $1S2M$, while the Twiss parameters propagation method is labeled $2S0M$.
The modified versions are labeled $2S3M$ and $3S0M$, respectively.


We define one normalized observable quantity for each error localization method.
For each value of this quantity, the specific sections and monitors used in this value calculation are identified.
For each section or monitor, distinct error indicators are constructed by aggregating all normalized observable values associated with that specific section or monitor.
Following summation, distinct normalized indicators for both sections and monitors are obtained.
The overall effectiveness of error localization is enhanced by combining these indicators obtained with different localization methods, thereby leveraging the joined strengths of each approach.


\begin{figure*}[th!]
    \begin{center}
    \minipage{0.5\linewidth}
      \includegraphics[width=\columnwidth,height=0.5\columnwidth]{fig/04_difference_action_and_phase.png}
    \endminipage
    \hfill
    \minipage{0.5\linewidth}
      \includegraphics[width=\columnwidth,height=0.5\columnwidth]{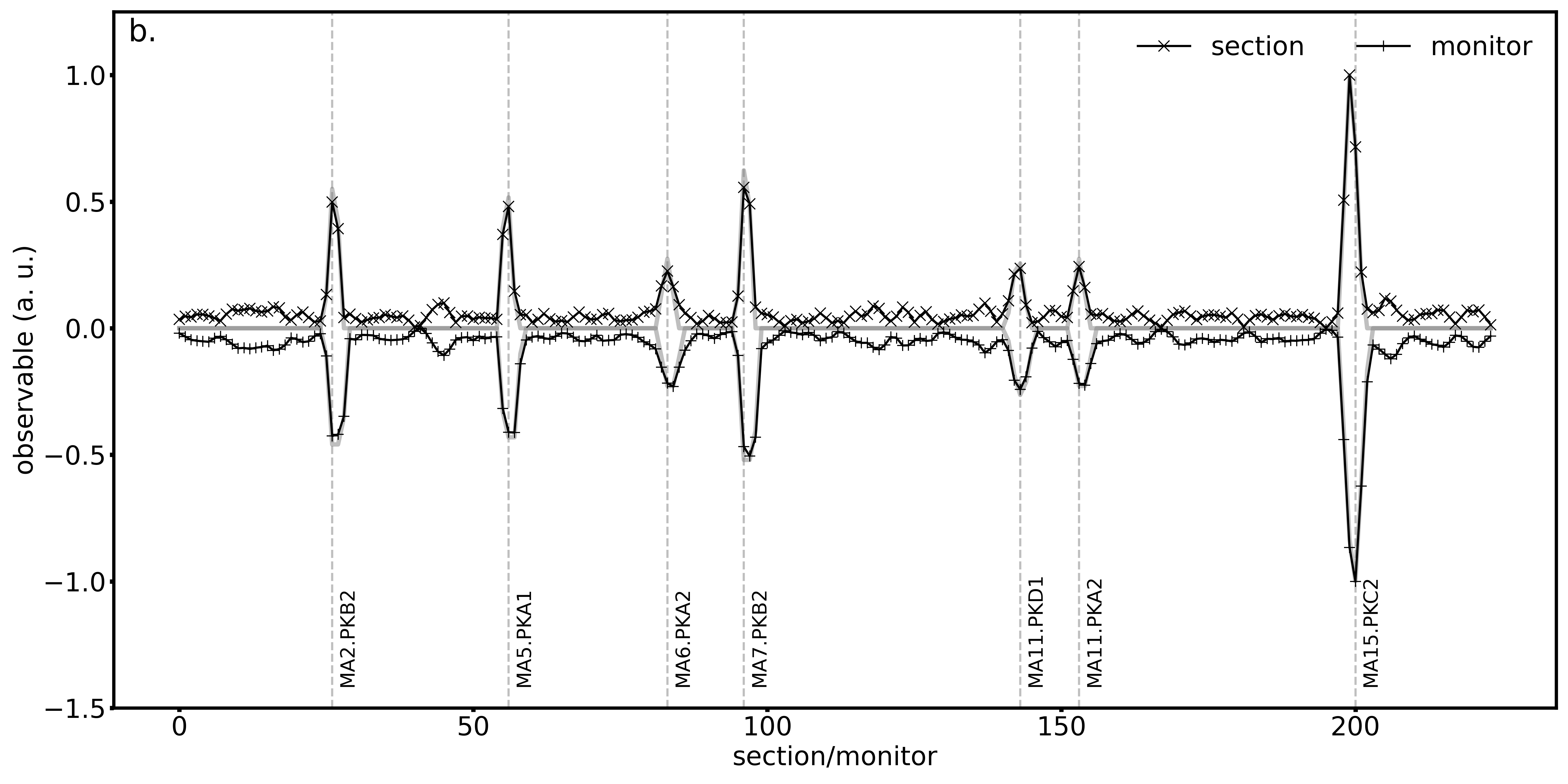}
    \endminipage
    \end{center}
    \caption{
    Error indicators based on the difference action and phase jump $2S3M$ (a.) and on the comparison of transverse momenta $2S3M$ (b.). Results for monitors are reflected relative to the horizontal axis. Error indicators without weak errors are shown in gray.
    }
    \label{fig:momenta}
\end{figure*}


Figure~\ref{fig:mod} shows error indicators for the difference formulations of the action and phase jump method, as well as the Twiss parameters propagation method.
In the action and phase jump method, the values of $I$ and $\varphi$ for transverse coordinates are first normalized independently, then added together, and normalized again. 
From the results for horizontal and vertical planes, a joined error indicator for sections and monitors is constructed.
In the Twiss parameters propagation method, the norm of the normalization matrix is used instead of individual Twiss parameters.
After normalizing this observable, the matrix norm values are used to construct the error indicator for sections.
An error indicator for monitors is not constructed, since the method is independent on monitor calibration errors.
Figure~\ref{fig:mod} shows that error locations are highlighted with peaks.
However, the localization accuracy for the difference methods deteriorates due to the use of more sections and monitors.
Like the original methods, the difference methods were able to successfully identify most of the strong errors.


\section{Comparison of Reconstructed Momenta}
\label{sec:momenta}


Transverse beam centroid momenta can be computed using coordinates from two BPMs and the model transport mapping between them~\cite{berz, ipac2016, ndmap}.
By comparing transverse momenta at the same monitor, obtained from different sections, an error localization method can be constructed.
This approach is not limited to linear mappings~\cite{ipac2016}.
This enables the localization of nonlinear focusing errors, although in this paper, it is used solely for localizing linear focusing errors.


Using coordinates from two BPMs $i$ and $j$, corresponding transverse momenta can be uniquely computed in a linear approximation. 
For example, the momenta at monitor $i$ can be expressed using coordinates $(q_{x, i}, q_{y, i}, q_{x, j}, q_{y, j})$ and the transport matrix $m^{(i,j)}$, which can include transverse coupling, between monitors as follows
\begin{widetext}
\begin{align}
p_{x, i}  &= q_{x,i}(m^{(i,j)}_{1,1}  m^{(i,j)}_{3,4} - m^{(i,j)}_{1,4}   m^{(i,j)}_{3,1})/(m^{(i,j)}_{1,4} m^{(i,j)}_{3,2} - m^{(i,j)}_{1,2} m^{(i,j)}_{3,4})
    + q_{x,j} m^{(i,j)}_{3,4}/(m^{(i,j)}_{1,2}   m^{(i,j)}_{3,4} - m^{(i,j)}_{1,4}   m^{(i,j)}_{3,2})  \nonumber \\
    &+ q_{y,i}(m^{(i,j)}_{1,3}  m^{(i,j)}_{3,4} - m^{(i,j)}_{1,4}   m^{(i,j)}_{3,3})/(m^{(i,j)}_{1,4} m^{(i,j)}_{3,2} - m^{(i,j)}_{1,2} m^{(i,j)}_{3,4})
    + q_{y,j} m^{(i,j)}_{1,4}/(m^{(i,j)}_{1,4}   m^{(i,j)}_{3,2} - m^{(i,j)}_{1,2}   m^{(i,j)}_{3,4}) 
\end{align}
\begin{align}
p_{y,i}  &= q_{x,i} (m^{(i,j)}_{1,1}  m^{(i,j)}_{3,2} - m^{(i,j)}_{1,2}   m^{(i,j)}_{3,1})/(m^{(i,j)}_{1,2} m^{(i,j)}_{3,4} - m^{(i,j)}_{1,4} m^{(i,j)}_{3,2})
    + q_{x,j}  m^{(i,j)}_{3,2}/(m^{(i,j)}_{1,4}   m^{(i,j)}_{3,2} - m^{(i,j)}_{1,2}   m^{(i,j)}_{3,4})\nonumber \\
    &+ q_{y,i} (m^{(i,j)}_{1,2}  m^{(i,j)}_{3,3} - m^{(i,j)}_{1,3}   m^{(i,j)}_{3,2})/(m^{(i,j)}_{1,4} m^{(i,j)}_{3,2} - m^{(i,j)}_{1,2} m^{(i,j)}_{3,4})
    + q_{y,j}  m^{(i,j)}_{1,2}/(m^{(i,j)}_{1,2}   m^{(i,j)}_{3,4} - m^{(i,j)}_{1,4}   m^{(i,j)}_{3,2}) 
\end{align}
\end{widetext}


Transverse momenta for any monitor can be computed using the coordinates from neighboring monitors.
The momenta values should match if the sections adjacent to the monitors and the monitors themselves do not contain errors.
The measure of momenta agreement is defined as the sum of the squares of the differences between momenta computed using different sections for both transverse planes
\begin{equation}
    \sum_n {\left(p^{n}_{x, i} - p^{n}_{x, j}\right)^2 + \left(p^{n}_{y, i} - p^{n}_{y, j}\right)^2}
\end{equation}
where summation over data samples is performed.


Data samples may be obtained from one or several TbT measurements.
If errors are present, this observable will stand out against the values at sections without errors.
Transverse momenta are computed using TbT data from BPMs.
TbT signals are filtered prior to computing momenta to reduce the effect of measurement noise.


\begin{figure*}[!t]
    \begin{center}
    \minipage{0.5\linewidth}
      \includegraphics[width=\columnwidth,height=0.5\columnwidth]{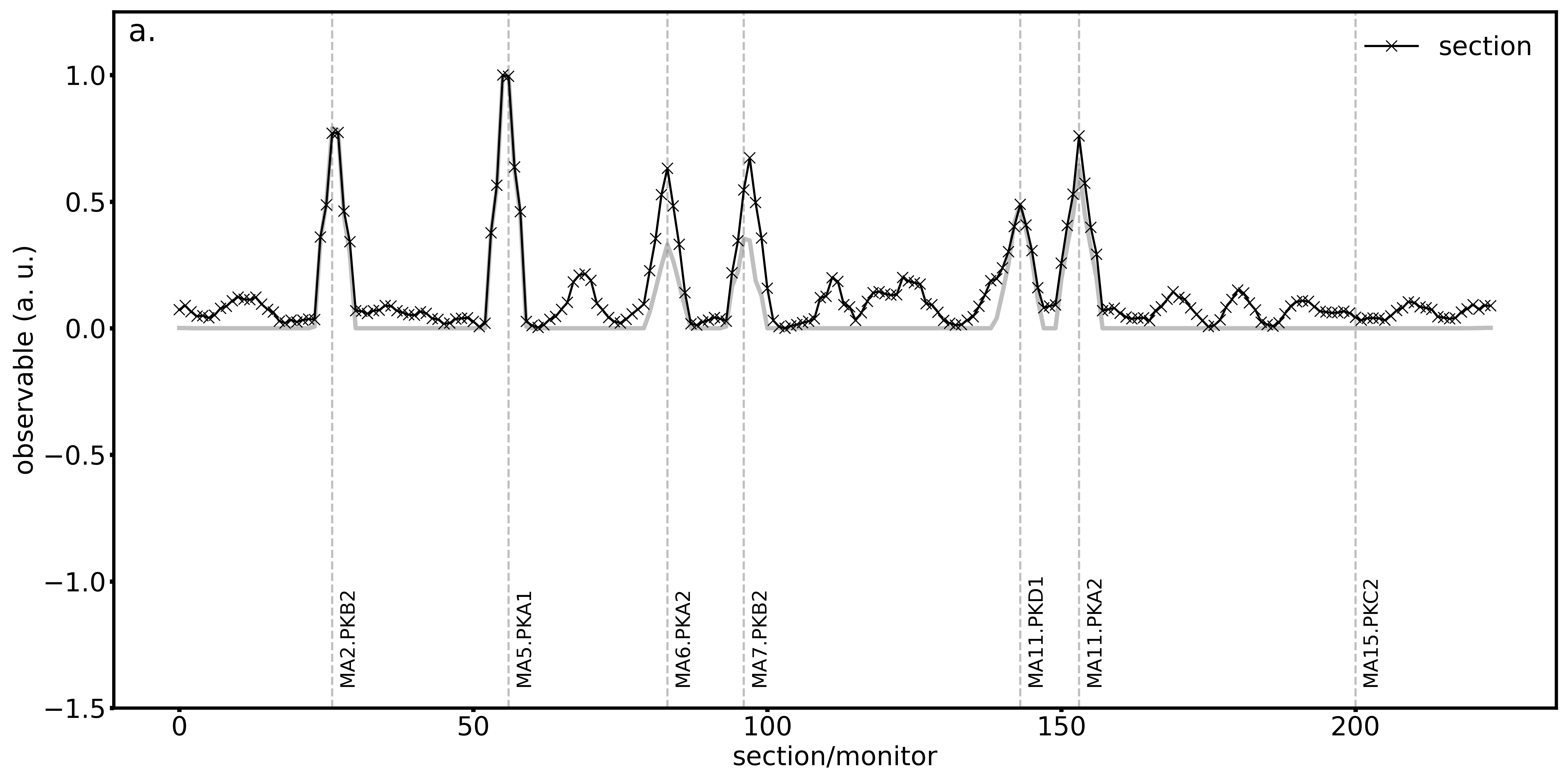}
    \endminipage
    \hfill
    \minipage{0.5\linewidth}
      \includegraphics[width=\columnwidth,height=0.5\columnwidth]{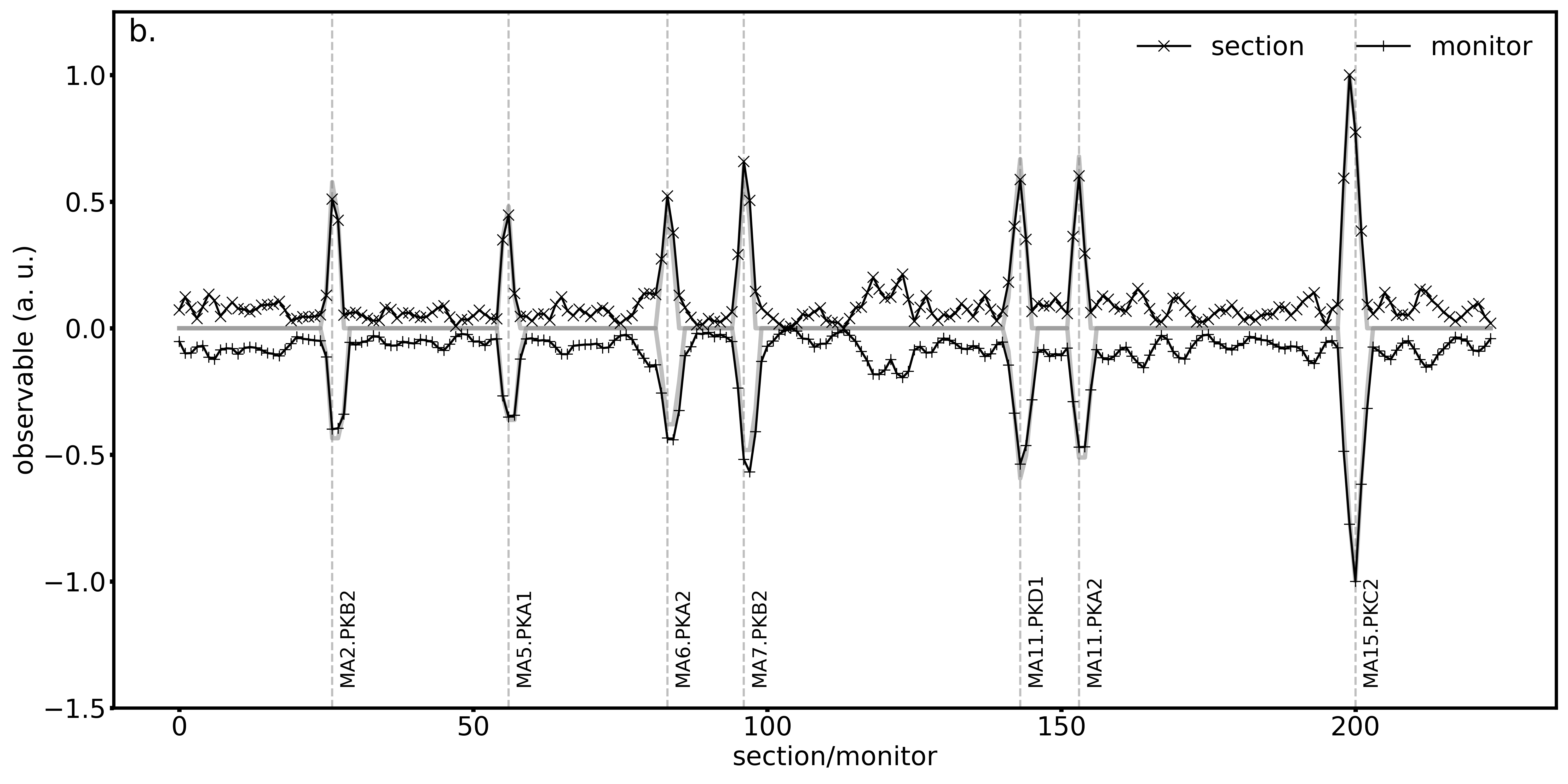}
    \endminipage
    \end{center}
    \caption{
    Error indicators based on the comparison of Twiss parameters, estimated from phase $4S0M$ (a.) and from approximation of linear invariants $2S3M$ (b.). Results for monitors are reflected relative to the horizontal axis. Error indicators without weak errors are shown in gray.
    }
    \label{fig:lrtwiss}
\end{figure*}


Figure~\ref{fig:momenta} shows error indicators for sections and monitors, which are constructed based on the comparison of transverse momenta.
All the introduced strong errors, including coupling errors, are clearly highlighted.
In this method, coupling errors are more pronounced, while some normal focusing errors are less visible.
This method is also sensitive to monitor calibration errors and has the same resolution as the difference action and phase jump method.


\section{Comparison of Twiss Parameters}
\label{sec:twiss}


Twiss parameters for uncoupled transverse motion can be calculated using phase measurements from three BPMs along with the model transport matrices between them.
This Twiss estimation method, as already noted, does not depend on monitor calibrations.
Given a probed monitor, Twiss parameters at its location can be determined using either two monitors to its left or two monitors to its right.
This yields two sets of Twiss parameters, which should match if the used sections contain no errors.
In this error localization method, the observable quantity is the norm of the difference between the left and right normalization matrices.
To compute a single value of this observable quantity at a probed monitor, four sections should be utilized: two on the left and two on the right.


Twiss parameters in the coupled case can be estimated using two BPMs and the section between them~\cite{coupled_twiss}.
For example, transverse momenta can be computed from coordinates at two BPMs (as discussed in Section~\ref{sec:momenta}) and a coupled normalization matrix can be fitted using the complete set of transverse phase space coordinates.
The coupled Twiss parameters can be estimated by minimizing the following function
\begin{equation}
\label{equation:objective}
    \min_{N, I_x, I_y} {\left(Q_x^2 + P_x^2 - 2 I_x\right)}^2 {\left(Q_y^2 + P_y^2 - 2 I_y\right)}^2 
\end{equation}
where $(Q_x,P_x,Q_y,P_y) = N^{-1} (q_x,p_x,q_y,p_y)$ are normalized coordinates, which are constructed from TbT coordinates $(q_x,p_x,q_y,p_y)$ and normalization matrix $N$. 
The values of the linear invariants $I_x$ and $I_y$ are also estimated from the optimization. 
In this case, the left and right values of the normalization matrices $N_{+}$ and $N_{-}$ can be compared in a similar manner. 
This method requires two sections and three BPMs. 
Since coordinates are used directly to estimate coupled Twiss parameters, the method is sensitive to monitor calibration errors.


Error localization indicators for these methods are shown in Figure~\ref{fig:lrtwiss}. 
As expected, when Twiss parameters are calculated based on phase measurements, calibration errors in the monitors do not manifest.
Similar to the difference Twiss parameters propagation method, this method is sensitive to focusing errors in areas with large beta functions.
However, coupling errors are more distinctly highlighted.
The method that estimates Twiss parameters using reconstructed transverse momenta effectively highlights all significant errors, including monitor calibration error.


\section{Comparison of Linear Invariants}
\label{sec:invariant}


In the linear case, two invariants of transverse motion exist
\begin{equation}
    I_x =  \mathcal{N}^{-1} \frac{1}{2} \left( Q_x^2 + P_x^2 \right), \quad I_y =  \mathcal{N}^{-1} \frac{1}{2} \left( Q_y^2 + P_y^2 \right)
\end{equation}
where the transformation $\mathcal{N}^{-1}$ converts Floquet coordinates $(Q_x, P_x, Q_y, P_y)$ into laboratory coordinates.
Since the normalization transformation is linear, it is possible to use normalization matrix $(Q_x,P_x,Q_y,P_y) = N^{-1} (q_x,p_x,q_y,p_y)$ as in the Section~\ref{sec:twiss}. 
The values of these invariants can be used to construct error localization method. 
In this paper, the invariants are estimated using the amplitudes and phases of TbT signals, as well as the values obtained from minimization of the objective defined in equation~\ref{equation:objective}.


In the uncoupled case, the transverse coordinates at the BPM $m$ change according to
\begin{align}
  q_m(t) &= a_m \cos(2 \pi \nu t + \varphi_m) \nonumber \\
         &= \sqrt{2 I \beta_m} \cos(2 \pi \nu t + \varphi_m)    
\end{align}
where $a_m$ and $\varphi_m$ are amplitude and phase of oscillations, $\nu$ is the frequency, $I$ is the linear invariant value, and $\beta_m$ is the beta function value.


\begin{figure*}[th!]
    \begin{center}
    \minipage{0.5\linewidth}
      \includegraphics[width=\columnwidth,height=0.5\columnwidth]{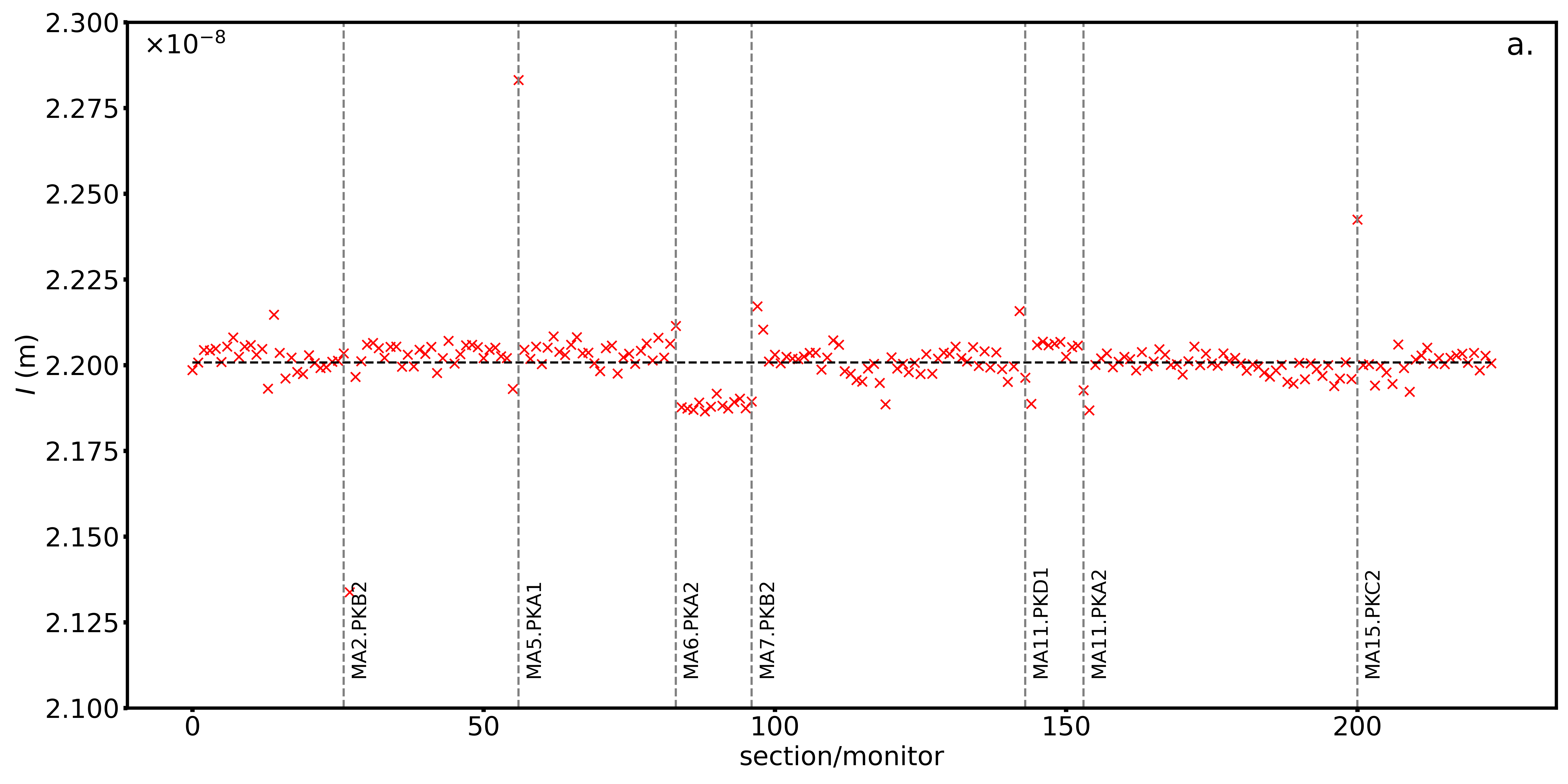}
    \endminipage\hfill
    \minipage{0.5\linewidth}
      \includegraphics[width=\columnwidth,height=0.5\columnwidth]{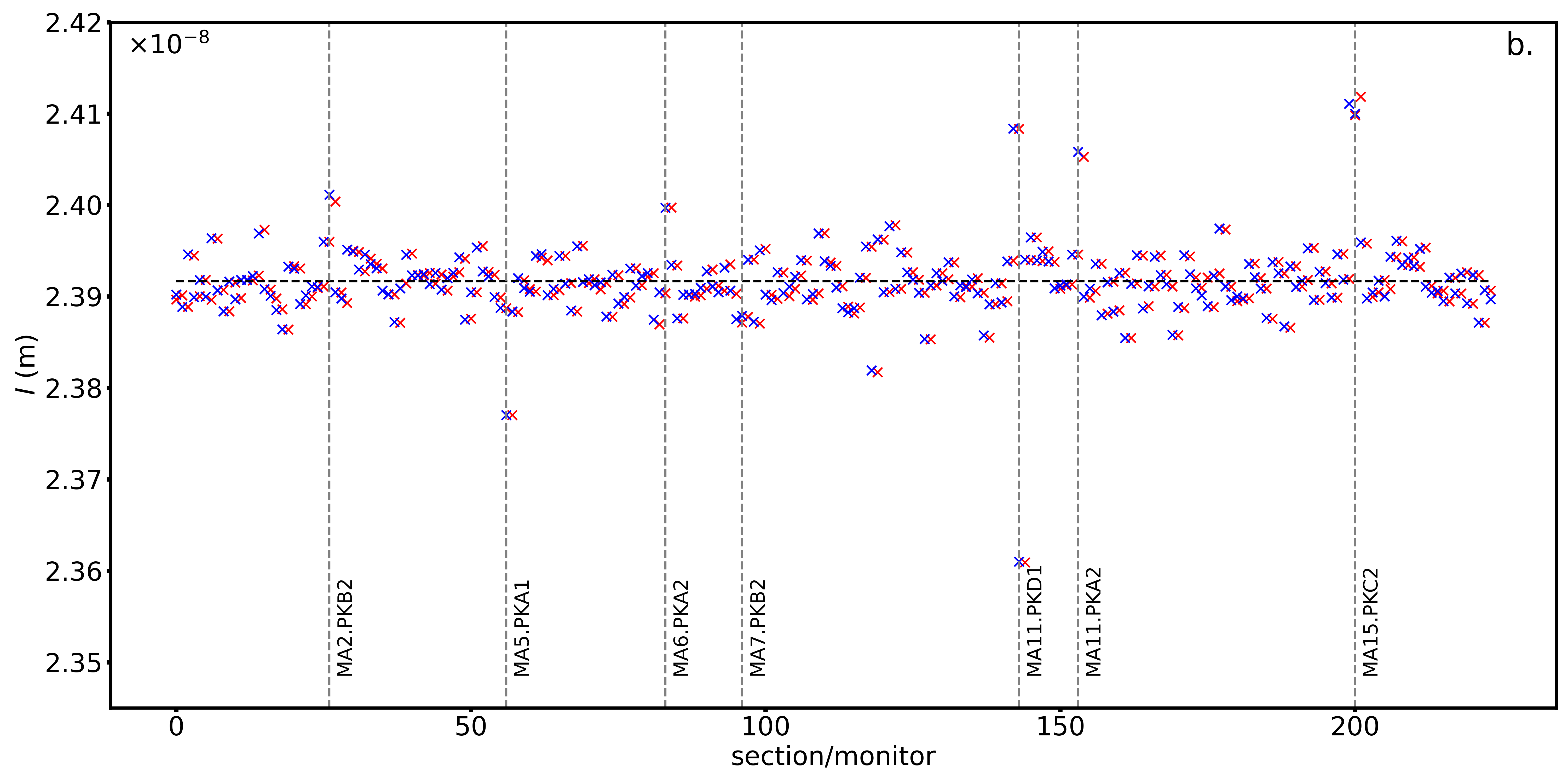}
    \endminipage
    \end{center}
    \caption{Estimated uncoupled (a.) and coupled (b.) linear invariants. For the latter case, both left (in red) and right (in blue) invariants are shown.}
    \label{fig:inv}
\end{figure*}


The uncoupled invariant value at the probed monitor can be computed using the amplitude and the beta function value.
The amplitude can be obtained from TbT signals, and the beta function value is estimated from phase using two additional BPMs around the probed one. 
$I = a_m^2 / \beta_m$ gives an estimate of the uncoupled invariant in the uncoupled case. 
Figure~\ref{fig:inv} shows the estimated invariant values for one transverse plane. 
This observable has peaks at the monitor locations for which monitors or sections with errors are used.
Additionally, jumps are observed at the locations of coupling errors.
Peaks are also appear due to monitor calibration errors, since the amplitude is used to estimate uncoupled invariants.


Figure~\ref{fig:inv} also shows estimates of coupled linear invariants obtained through minimization.
Determining the normalizing matrix and invariants requires at least two BPMs, with the second monitor being utilized to reconstruct transverse momenta at the first one.
Within a given section, invariants can be estimated using either the left or right monitors.
Consequently, one can obtain corresponding left and right coupled invariants.
Similar to the uncoupled case, peaks indicate the locations of errors.
Coupling errors are also manifested as peaks.


\begin{figure}[ht!]
    \begin{center}
    \includegraphics[width=\columnwidth,height=0.5\columnwidth]{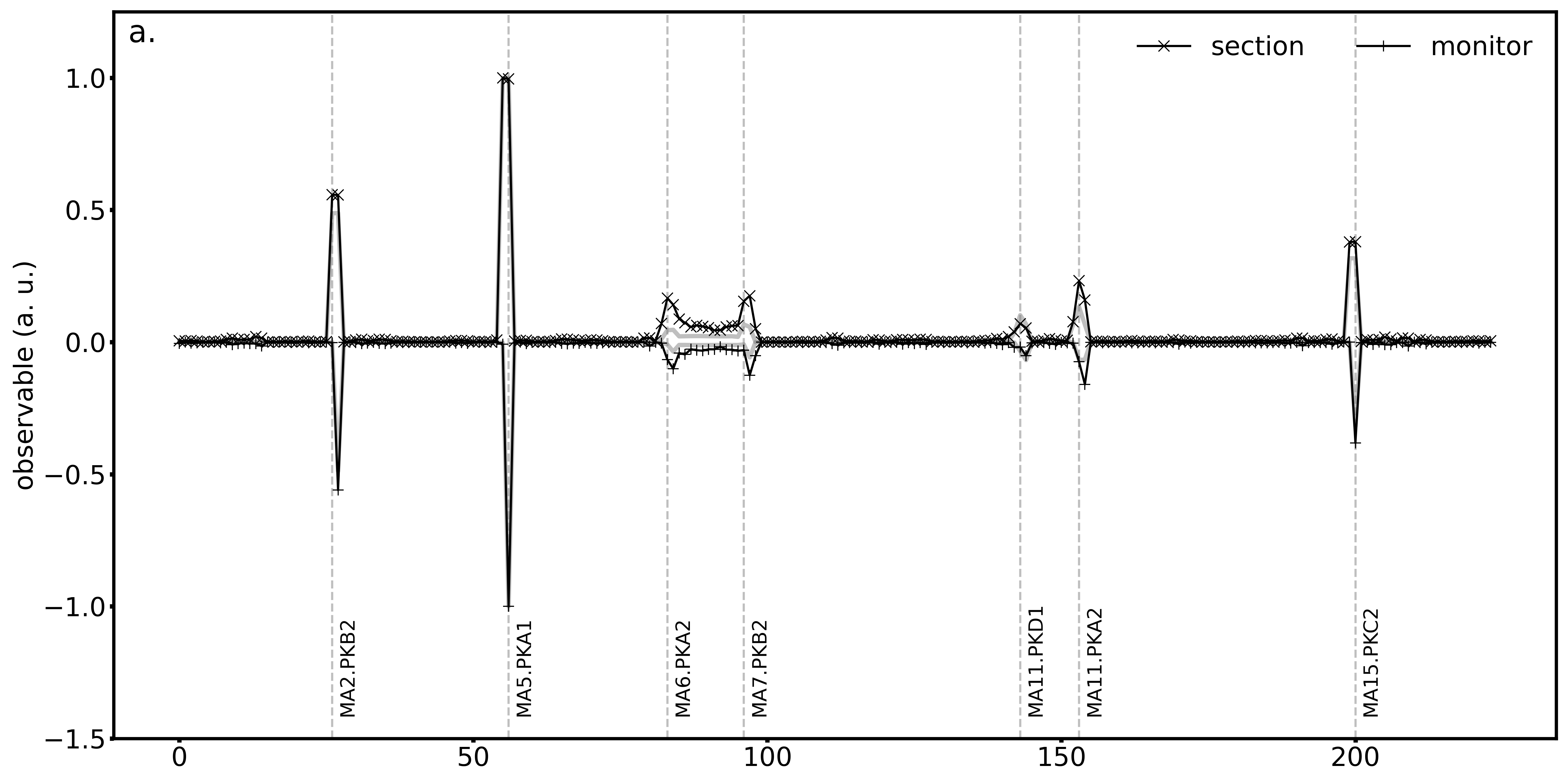}
    \includegraphics[width=\columnwidth,height=0.5\columnwidth]{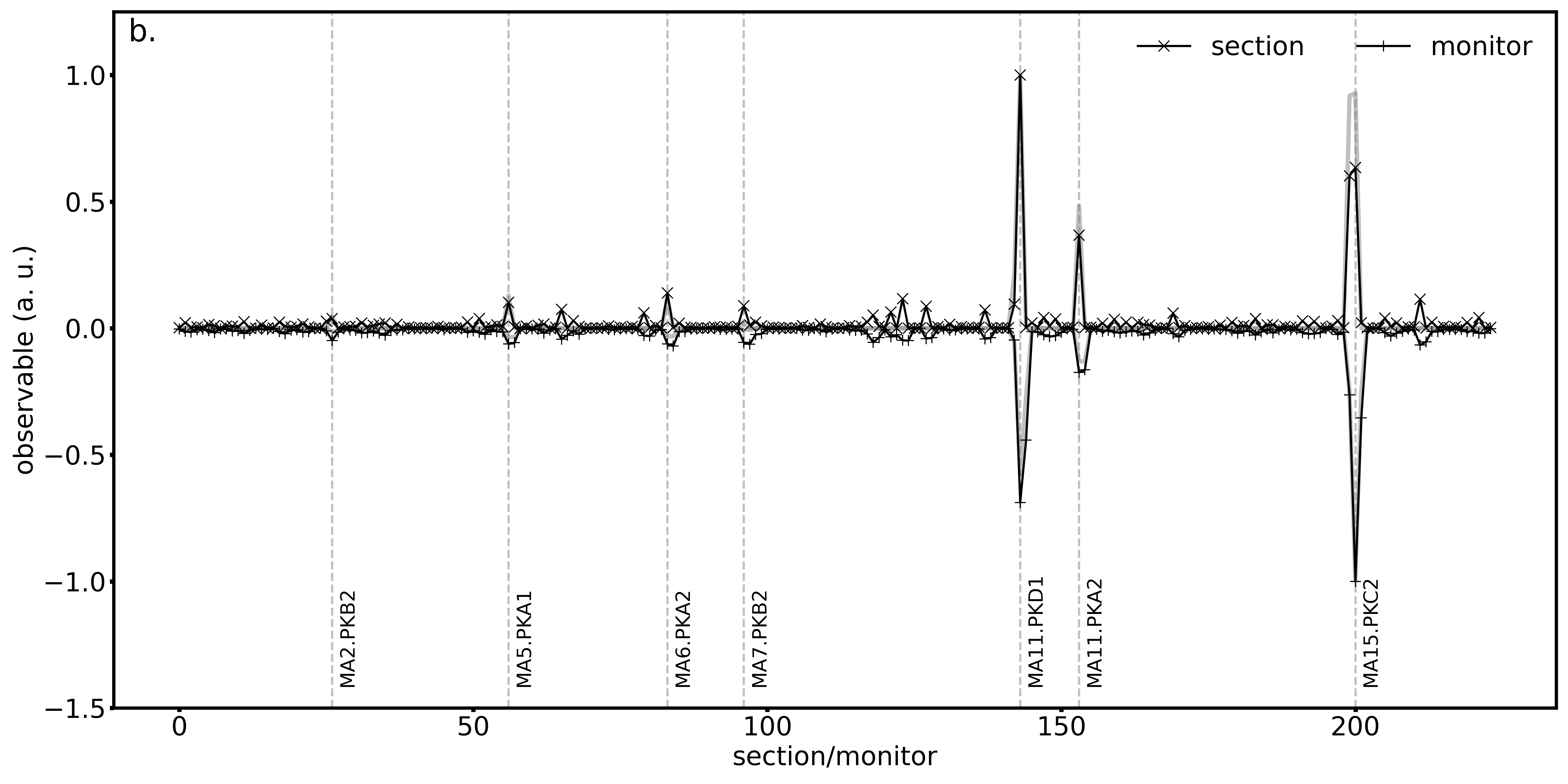}
    \end{center}
    \caption{Error indicators based on uncoupled $2S1M$ (a.) and coupled linear $1S2M$ (b.) invariants. Results for monitors are reflected relative to the horizontal axis. Error indicators without weak errors are shown in gray.}
    \label{fig:invs}
\end{figure}


Error localization indicators can be constructed based on the estimated invariants.
First, the invariant values for all points are reflected relative to the median value. 
These adjusted values are then normalized and used as observables for constructing the indicators.
Figure~\ref{fig:invs} shows the results of this procedure when applied to both methods of invariant estimation.
In the uncoupled case, the normalized observables for horizontal and vertical planes are multiplied together.
This is done to reduce the effects of jumps from coupling errors.
Although jumps can be eliminated using differences, this approach would reduce the resolution of localization.
Most strong errors are correctly highlighted, and this indicator has the best possible resolution for localizing monitor calibration errors.


In the coupled case, strong errors are similarly highlighted.
The observables for left and right invariants are multiplied.
This latter method has the same resolution as the original action and phase jump method.
The difference between left and right coupled invariants appears to be sensitive to coupling errors.
In the absence of weak errors, all significant coupling errors are distinctly highlighted, and normal focusing errors are less noticeable.
However, in the presence of weak background errors, the reliability of this difference becomes unreliable.


\section{Comparison of Transport Matrices}
\label{sec:matrix}


\begin{figure*}[th!]
    \begin{center}
    \minipage{0.5\linewidth}
      \includegraphics[width=\columnwidth,height=0.5\columnwidth]{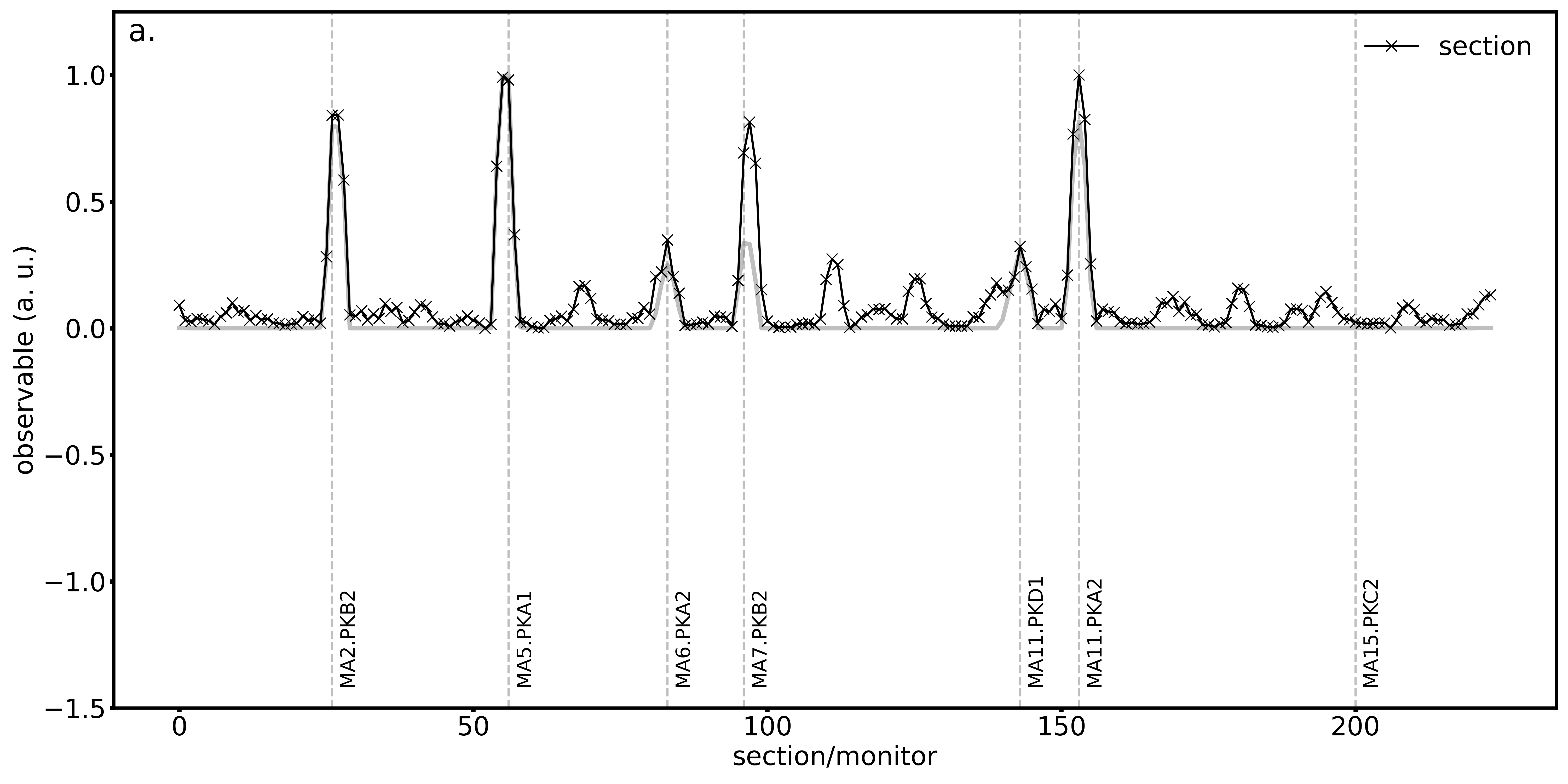}
    \endminipage\hfill
    \minipage{0.5\linewidth}
      \includegraphics[width=\columnwidth,height=0.5\columnwidth]{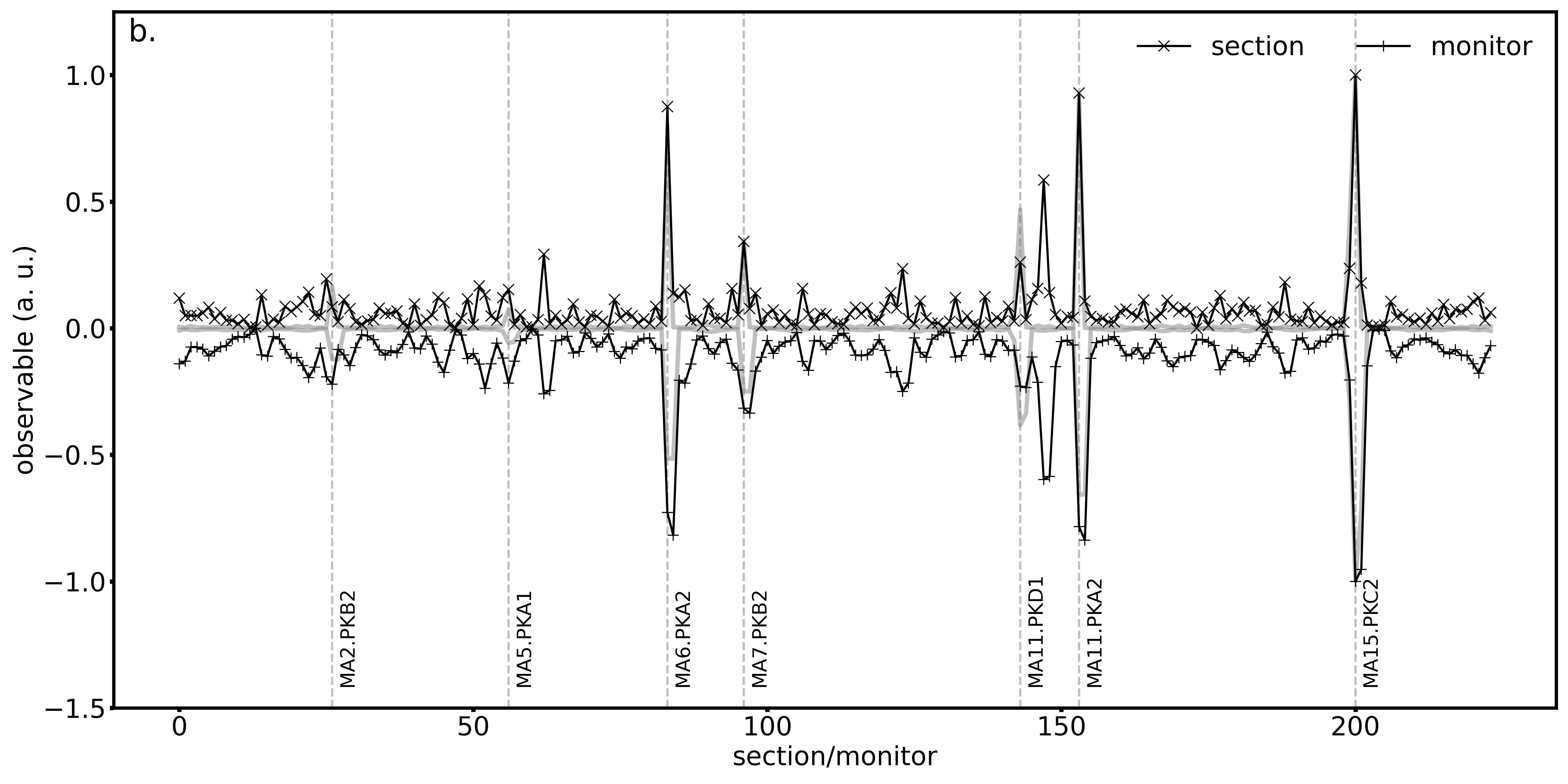}
    \endminipage
    \end{center}
    \caption{Error indicators based on transport matrices computed using phase data  $3S0M$ (a.) and linear invariants fit $1S2M$ (b.). Results for monitors are reflected relative to the horizontal axis. Error indicators without weak errors are shown in gray.}
    \label{fig:matrix}
\end{figure*}


The transport matrix between adjacent BPMs can be calculated if the Twiss parameters at the monitors and the phase advances between them are known
\begin{equation}
    \label{equation:transport}
    M_{i, j} = N_{j} R_{i, j}(\varphi_{i, j}) N_{i}^{-1}
\end{equation}
where $R_{i, j}(\varphi_{i, j})$ is the rotation matrix with angle $\varphi_{i, j}$, $N_{i}$ and $N_{j}$ are normalization matrices corresponding to monitors $i$ and $j$, respectively.
The phase advance $\varphi_{i, j}$ is computed using the measured phases at monitors $i$ and $j$.
Phase advances can be computed from TbT signals.
For estimating Twiss parameters, the two methods described previously in Section~\ref{sec:twiss} can be used.
These estimated transport matrices are then compared with the model transport matrices.
The norm of the difference between the estimated and the model transport matrix is used as an observable quantity.
The results of using such norm errors for construction of error localization indicators are shown in Figure~\ref{fig:matrix}. 
To determine transport matrices in the uncoupled case using phase data, data from three sections between the monitors are used.
In the coupled case, one section and two BPMs are used, which is analogous to the action and phase jump method.


In both methods, the majority of strong errors are effectively highlighted.
As expected, when Twiss parameters are calculated using phase data, the method is not affected by monitor calibration errors, but it is sensitive to weak errors at locations with large beta function values.
This method  also has the highest resolution among phase based methods, where errors manifest as peaks.
When coupled Twiss parameters are used, the results become sensitive to monitor calibration errors.
Additionally, the skew block norm of the transport matrix can be used to identify skew focusing errors.
However, similar to the difference in left and right coupled linear invariants, this indicator becomes unreliable in the presence of weak errors.
It is also worth mentioning, given $N_{i}$ and $M_{i, j}$, it is possible to compute  $\varphi_{i, j}$~\cite{wolski}.
This enables the construction of an indicator based on the comparison of model phase advance with that obtained from the measured normalization matrix and model transport matrix, but its results are similar to the indicator based on transport matrices.


\section{Combining Error Localization Indicators}
\label{sec:combine}


The localization methods discussed vary in their sensitivity to different types of errors.
Therefore, combining these methods could improve the overall accuracy of error localization.
Moreover, these methods have different localization resolutions for the sections between monitors as well as for the monitors themselves.
Methods with the highest resolution can be used along with combined results.
As mentioned, each error localization method yields two normalized error localization indicators: one for sections and one for monitors.
The results for each case can be summed to enhance overall sensitivity, or multiplied to emphasize locations where the methods are triggered simultaneously.

\begin{figure*}[!t]
    \begin{center}
    \minipage{0.5\linewidth}
      \includegraphics[width=\columnwidth,height=0.5\columnwidth]{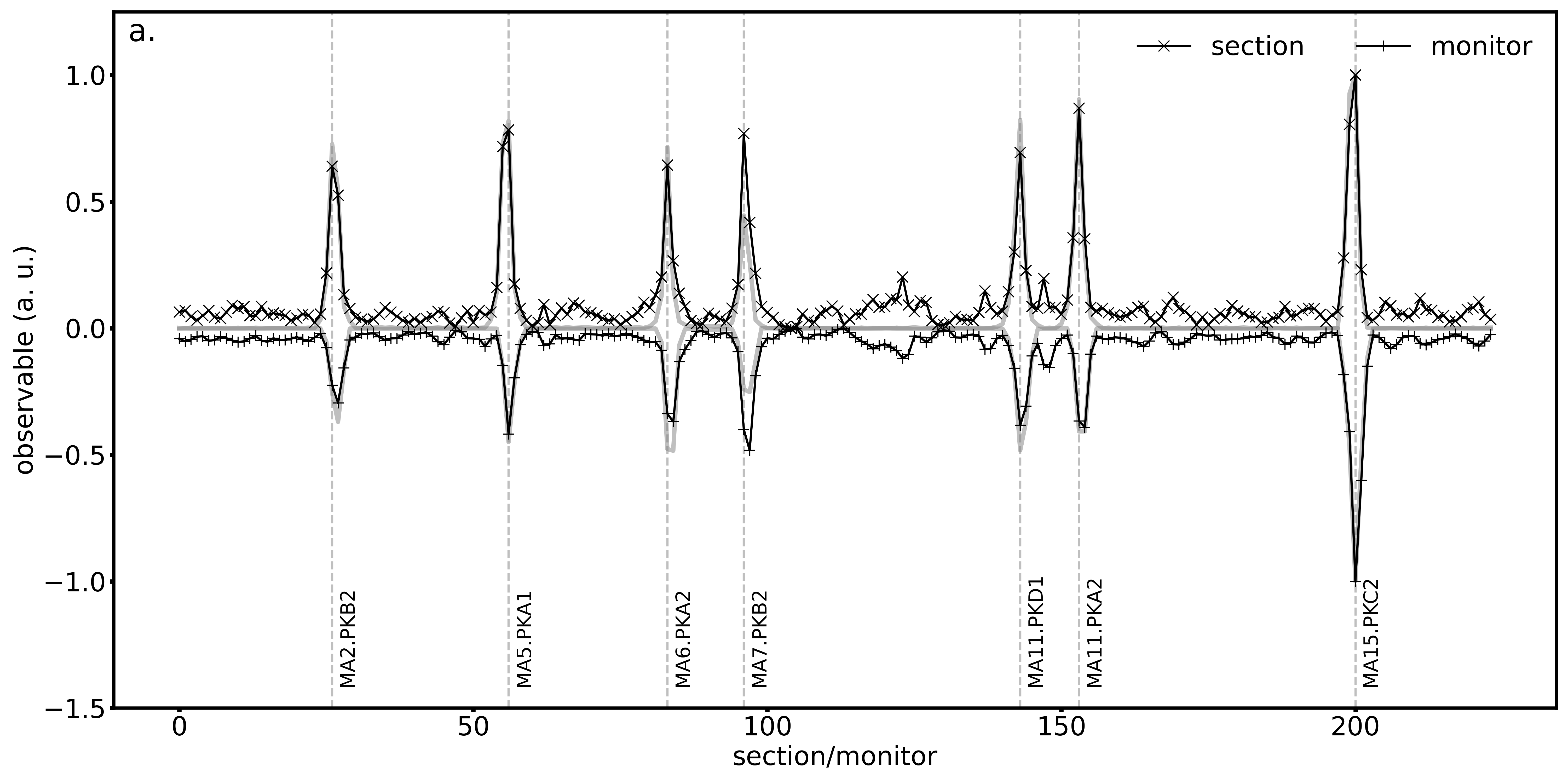}
    \endminipage
    \hfill
    \minipage{0.5\linewidth}
      \includegraphics[width=\columnwidth,height=0.5\columnwidth]{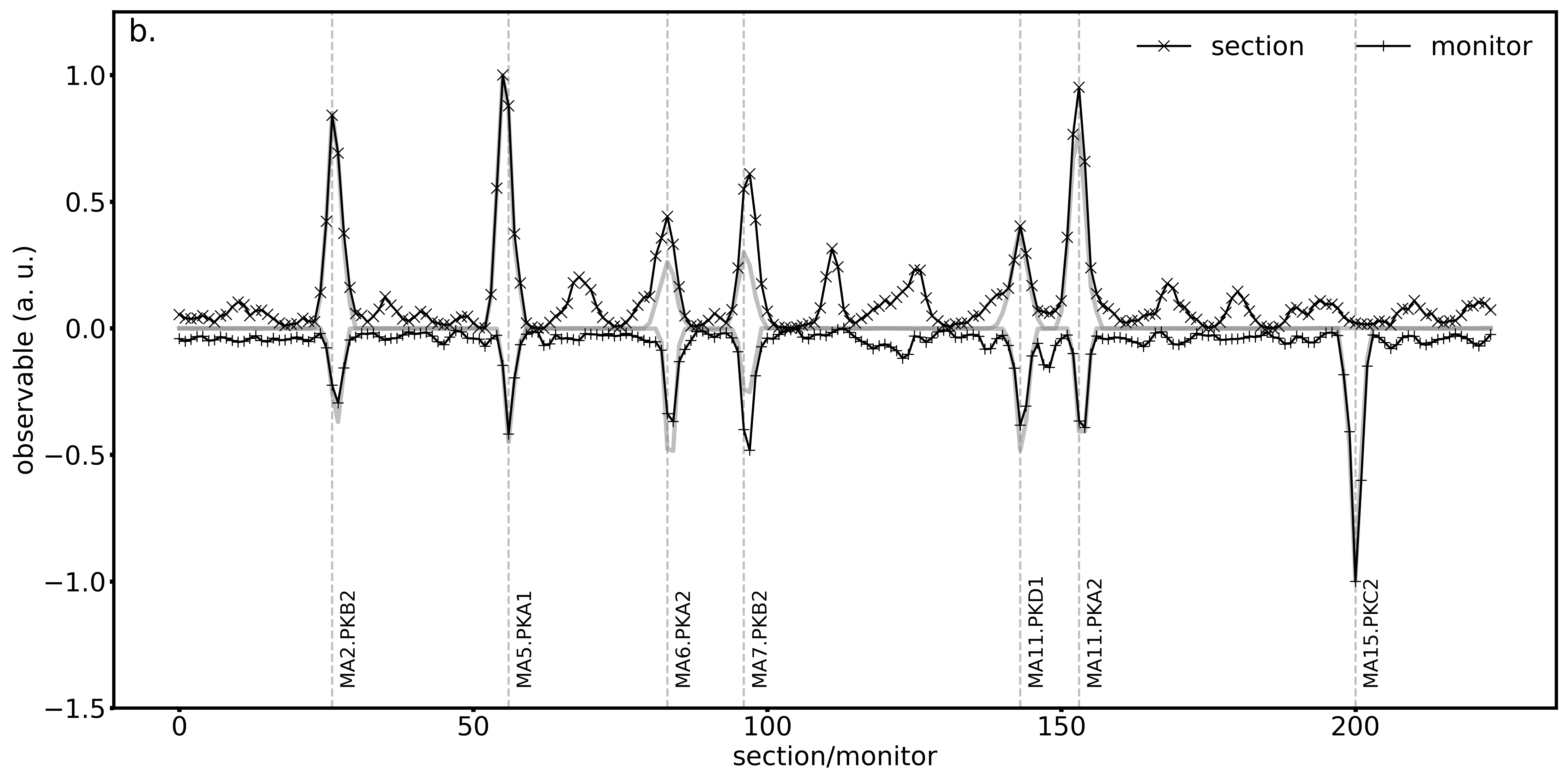}
    \endminipage
    \end{center}
    \caption{Combination of all error indicators (a.), combination of methods based only on phase data (b.). For comparison, the overall result for monitors is also shown in (b.).}
    \label{fig:all}
\end{figure*}

The method that compares reconstructed momenta and the difference action and phase jump method both exhibit identical localization resolutions.
The results of these methods are combined and then normalized.
The results of all other error localization methods are then added to this intermediate result.
Figure~\ref{fig:all} shows the combined result of all these methods.
All strong errors are clearly pinpointed in the combined result.
The maximum of each peak accurately aligns with the corresponding section or monitor where a strong focusing error was introduced.
Figure~\ref{fig:all} also shows the result of combining methods based only on phase data. 
This combination can be compared with the overall result for monitors to identify potential monitor calibration errors.
For the test problem, this comparison effectively identifies the calibration error.
In addition, the method based on comparing coupled linear invariants can be used, since it has the best possible resolution for monitor calibration errors.
Methods that are sensitive to skew focusing errors can be compared with the overall results for sections to locate these specific errors.


In the test problem, each peak corresponds to sections or monitors with errors.
However, this is not assured in more general cases.
The error could be located near the peak maximum, thus it is necessary to select of not just the peak's maximum location but also the surrounding area.
Additionally, the method of comparing transport matrices can be utilized for verification purposes, as it has the best possible resolution for error localization in the sections between BPMs.


\section{Summary}
\label{sec:summary}


Several methods for the localization of linear focusing errors have been explored.
To facilitate convenient automatic error localization used to reduce parameter space for efficient correction, all methods have been formulated to identify potential focusing errors as peaks in certain observable quantities.
Two distinct error localization indicators are introduced to differentiate between errors in the sections between monitors and calibration errors of the monitors.
Common error localization methods have been reviewed and modified into a difference form to transition from jumps to peaks analysis.
Several new error localization methods have been developed based on the comparison of reconstructed momenta, Twiss parameters, both uncoupled and coupled linear invariants, and transport matrices.
All described methods can use TbT signals from BPMs. 
The main idea of these methods is to calculate an observable quantity by using different sections between monitors.
For the test problem, these methods have successfully demonstrated their capability to detect strong errors that were introduced.


The methods have different sensitivities and resolutions, and their combination improves the overall accuracy. 
Furthermore, a combination of methods that are not sensitive to monitor calibration errors can be compared with the total combination to identify potential calibration errors.
The method based on estimation of uncoupled linear invariants has the best possible resolution for a monitor error indicator.
In contrast, the method based on transport matrix from coupled Twiss parameters has the best possible resolution for a section error indicator.


\section{Acknowledgments}

This work was partially supported by the Ministry of Science and
Higher Education of the Russian Federation within the governmental order for
Boreskov institute of Catalysis (project FWUR-2022-0001).
 


%

\end{document}